\documentclass[15pt a4paper]{article}

\usepackage{latexsym}
\usepackage{amsmath}
\usepackage{amssymb}
\usepackage{graphicx}

\graphicspath{{./}{./figs/}}

\begin{document}

\title{Hopf Bifurcation and Chaos in Tabu Learning Neuron Models \thanks{Accepted by International Journal of Bifurcation and Chaos}}
\author{Chunguang Li$^1$\thanks{Corresponding author, Email:
cgli@uestc.edu.cn}, Guanrong Chen$^2$, Xiaofeng Liao$^1$, Juebang
Yu$^1$}

\date{\small $^1$Institute of Electronic Systems, School of Electronic
Engineering, \\University of Electronic Science and
Technology of China,\\ Chengdu, 610054, P. R. China.\\
$^2$Department of Electronic Engineering, City University of Hong
Kong, \\83 Tat Chee Avenue, Kowloon, Hong Kong, P. R. China.}
\maketitle

\begin{abstract}
In this paper, we consider the nonlinear dynamical behaviors of
some tabu leaning neuron models. We first consider a tabu learning
single neuron model. By choosing the memory decay rate as a
bifurcation parameter, we prove that Hopf bifurcation occurs in
the neuron. The stability of the bifurcating periodic solutions
and the direction of the Hopf bifurcation are determined by
applying the normal form theory. We give a numerical example to
verify the theoretical analysis. Then, we demonstrate the chaotic
behavior in such a neuron with sinusoidal external input, via
computer simulations. Finally, we study the chaotic behaviors in
tabu learning two-neuron models, with linear and quadratic
proximity functions respectively. \medskip

{\bf Keywords:} Tabu learning, neural network, Hopf bifurcation,
chaos
\end{abstract}

\section{Introduction}

Neurons as the fundamental elements of the brain can generate
complex dynamical behaviors. The studies of nonlinear dynamics in
both biological and artificial neural network models are very
important in that, on the one hand, the studies may provide
explanations for the rich dynamics including oscillations and
chaos in biological neural networks, and on the other hand, if we
understand more about the dynamical behaviors of various neural
networks then we can design some control methods to achieve
desirable system behaviors in artificial neural networks.

In recent years, dynamical characteristics of various neural
networks have become a focal subject of intensive research
studies. Bifurcation and chaos have been identified and
investigated in many kinds of neural networks. For example,
chaotic solutions were obtained in [Kurten \& Clark, 1986], from a
neural network consisting of 26 neurons. In [Babcock \&
Westervelt, 1987], a two-neuron model was studied, with or without
time delays. Numerical solutions of differential equations with
electronic circuit models of chaotic neural networks were
qualitatively compared in [Kepler et al., 1990]. In [Das II et
al., 1991], a chaotic neural network with four neurons was
investigated, and chaotic behavior was found in [Zou \& Nossek,
1993] from a cellular neural network with three cells. In [Li et
al., 2001], we also presented some findings of chaotic phenomenon
in a three-neuron hysteretic Hopfield-type neural network. In
[Bondarenko, 1997], a high-dimensional chaotic neural network
under external sinusoidal force was studied. In [Ueta \& Chen,
2001], bifurcation and chaos as well as their control in a system
of strongly connected neural oscillators were discussed. In [Das
et al., 2002], chaos in a three-dimensional general neural network
model was investigated. In [Olien \& Belair, 1997; Wei \& Ruan,
1999; Gopalsamy et al., 1998; Liao et al., 1999; Liao et al.,
2001; Liao et al., 2001; Gilli, 1993; Lu, 2002], the bifurcation
and chaotic behaviors of various delayed neural networks were
studied. Moreover, the chaotic phenomenon in a
neural-network-based learning algorithm was reported in [Li et
al., 2003].

Tabu learning [Beyer \& Ogier, 1991] applies the concept of tabu
search [Glover, 1989; Glover, 1990] to neural networks for solving
optimization problems. By continuously increasing the energy
surface in a neighborhood of the current network state, it
penalizes those states that have already been visited. This
enables the state trajectory to climb out of local minima while
tending toward those areas that have not yet been visited, thus
performing an efficient search through the problem's solution
space. Note that, unlike most existing neural-network-based
methods for optimization, the goal of the tabu learning method is
not to force the network state to converge to an optimal or nearly
optimal solution, but rather, the network conducts an efficient
search through the solution space. Knowing this, it is very
natural for one to ask what the state trajectory of the tabu
learning neural network is like? In the present paper, the aim is
to provide an answer to this question by studying the dynamical
behaviors of the tabu learning neural network.

More precisely, this paper investigates the nonlinear dynamical
behaviors of a tabu leaning single neuron model and of two
two-neuron models, respectively. By choosing the memory decay rate
as a bifurcation parameter, it is proved that Hopf bifurcation
occurs in the tabu leaning single neuron model. The stability of
the bifurcating periodic solutions and the direction of the Hopf
bifurcation are then determined by applying the normal form
theory. Chaotic behavior of such a neuron with small sinusoidal
external input is also demonstrated by computer simulations. To
that end, the chaotic behaviors in tabu learning two-neuron models
with linear and quadratic proximity functions, respectively, are
studied by means of computer simulations.

The paper is organized as follows. The tabu learning neural
network is first described in Section 2. In Section 3, the Hopf
bifurcation of a single neuron model is studied. In Section 4, the
observed chaotic behavior of the single neuron model is described.
In Section 5, the dynamical behaviors of a two-neuron model with a
linear proximity function are studied. And, in Section 6, the
dynamical behaviors of a two-neuron model with a quadratic
proximity function are investigated. Finally, conclusions are
drawn in Section 7.

\section{Tabu learning neural network}

The common problem of minimizing an objective function in many
optimization schemes by using the Hopfield-type neural networks of
the form
\begin{equation}
C_i\dot{u}_i=-\frac{1}{R_i}u_i+\sum_jT_{ij}V_j+I_i
\end{equation}
can be translated into minimizing the following energy function:
\begin{equation}
E_0=-\frac{1}{2}\sum_i\sum_jT_{ij}V_iV_j-\sum_iI_iV_i
+\sum_i\frac{1}{R_i}\int_0^{V_i}f^{-1}(s)\,ds
\end{equation}
where $V_i=f(u_i)$, $f(\cdot)$ is the activation function, $u_i$
is the state of neuron $i$, $C_i$ and $R_i$ are positive
constants, $T_{ij}$ represents the strength of the connection from
neuron $j$ to neuron $i$, and $I_i$ represents the input current
to neuron $i$.

In the tabu leaning method, the energy surface $E_0$ is
continuously increasing in a neighborhood of the current state. At
time $t$, the energy surface is given by
\begin{equation}
E_t=E_0+F_t(V)
\end{equation}
with
\begin{equation*}
F_t(V)=\beta\int_0^te^{\alpha(s-t)}P(V,V(s))\,ds
\end{equation*}
where $\alpha$ and $\beta$ are positive constants and $P(V,W)$ is
a measure of the proximity of the two vectors $V$ and $W$ (thus,
$P(V,W)$ is maximized when $V=W$). Therefore, states ``nearest"
those already being visited are penalized the most, so that the
penalty encourages a search through states that have not yet been
visited. The exponential kernel keeps the integral from increasing
to infinity and results in a higher penalty for states being
visited most recently. The latter property also helps the network
in quickly climbing out of local minima.

In solving optimization problems, the memory decay rate $\alpha$
and the learning rate $\beta$ must be chosen carefully. If
$\alpha$ is too large, states already being visited are likely to
be revisited again; but if it is too small, the network may
require a long time to climb out of local minima. In addition,
$\beta$ must be chosen so as to achieve a balance between
searching for the states that minimize $E_0$ and searching for the
states that minimize $F_t(V)$.

One possible choice for the proximity function $P(V, W)$ is
defined by [Beyer \& Ogier, 1991]
\begin{equation}
P_1(V,W)=\sum_i(1+V_iW_i)
\end{equation}
where $V_i$ and $W_i$ are the components of vectors $V$ and $W$,
respectively. This is called the linear proximity function for it
is linear in $V$.

A weakness of $P_1(V,\,W)$ is that it penalizes too heavily those
vectors that are not close to $W$. A proximity function, which is
better in this respect, is given by [Beyer \& Ogier, 1991]
\begin{equation}
P_2(V,W)=\frac{1}{2}\sum_i\sum_{j\neq i}(1+V_iW_i)(1+V_jW_j)
\end{equation}
This is called the quadratic proximity function for it is
quadratic in $V$, which leads to a quadratic penalty term
$F_t(V)$.

If the linear proximity function is selected, then the neural
network that performs gradient descent on $E_t$ has the following
state equation [Beyer \& Ogier, 1991]:
\begin{equation}
\begin{array}{lcl}
C_i\dot{u}_i&=&-\frac{1}{R_i}u_i+\sum_jT_{ij}V_j+I_i
-\frac{\partial{F_t}(V)}{\partial{V_i}}\\
&=&-\frac{1}{R_i}u_i+\sum_jT_{ij}V_j+I_i+J_i(t)
\end{array}
\end{equation}
where
\begin{equation*}
J_i(t):= -\beta\int_0^te^{\alpha(s-t)}V_i(s)\,ds
\end{equation*}
Therefore, $J_i$ satisfy the following learning equation:
\begin{equation}
\dot{J_i}=-\alpha J_i-\beta V_i
\end{equation}

In the next two sections, a tabu learning single neuron model is
studied, which is described by
\begin{equation}
\begin{array}{l}
C\dot{x}=-x/R+af(x)+y+I \\
\dot{y}=-\alpha y-\beta f(x)
\end{array}
\end{equation}
where $C, R, a,\alpha$ and $\beta$ are all positive constants, and
$I$ is the external input.

\section{Hopf bifurcation in a tabu learning single neuron model}

\subsection{Existence of Hopf bifurcation}

In this subsection, conditions for the existence of a Hopf
bifurcation in a tabu learning single neuron model is derived. The
model is obtained by setting $C=1,\, R=1,\,\mbox{and}\, I=0$ in
system (8), as follows:
\begin{equation}
\begin{array}{l}
\dot{x}=-x+af(x)+y \\
\dot{y}=-\alpha y-\beta f(x)
\end{array}
\end{equation}

Suppose that in (9), $f\in C^3(R),\,f(0)=0$, and that $a,\alpha$
and $\beta$ satisfy the inequalities $a>0,\, b>0,\, \left|
{a-\frac{\beta}{\alpha}}\right|M<1$, where $|f^\prime(0)|<M$. It
is clear that system (9) have an unique equilibrium $(0,0)$ under
these conditions.

Expanding system (9) into first, second, third and other
higher-order terms about the equilibrium $(0,0)$, one has
\begin{equation}
\left[ \begin{array}{c} \dot{x}\\ \dot{y} \end{array}
\right]=\left[\begin{array}{cc}af^\prime(0)-1&1\\-\beta
f^\prime(0)&-\alpha\end{array}
\right]\left[\begin{array}{c}x\\y\end{array}\right]
+\left[\begin{array}{c}\frac{1}{2}af''(0)x^2
+\frac{1}{6}af'''(0)x^3+\cdots\\
-\frac{1}{2}\beta f''(0)x^2-\frac{1}{6}\beta f'''(0)x^3+\cdots
 \end{array}\right]
\end{equation}
The associated characteristic equation of its linearized system is
\begin{equation*}
\lambda^2+(1-af'(0)+\alpha)\lambda+(1-af'(0))\alpha+\beta f'(0)=0
\end{equation*}
Define
\begin{equation*}
b_1=b_1(\alpha)=(1-af'(0)+\alpha)
\end{equation*}
\begin{equation*}
b_2=b_2(\alpha)=(1-af'(0))\alpha+\beta f'(0)>0
\end{equation*}
Then, the Routh-Hurwitz criterion implies that the equilibrium
$(0,0)$ of system (9) is locally asymptotically stable if $b_1
(\alpha)>0$. If
\begin{equation*}
\alpha_0=af'(0)-1
\end{equation*}
then $b_1(\alpha_0)=0$, and the characteristic equation has one
pair of purely imaginary roots, $\lambda_{1,2}=\pm \omega_0i$,
where $\omega_0=\sqrt{b_2(\alpha_0)}=\sqrt{\beta f'(0)-
\alpha_0^2}$. It follows from simple calculation that
\begin{equation*}
\left.\frac{d[Re(\lambda_1)]}{d\alpha}\right|_{\alpha_0}
=-\frac{1}{2}<0
\end{equation*}

The above analysis is summarized as follows: \medskip

{\bf THEOREM 1:} If $\alpha>af^\prime(0)-1$, then the equilibrium
$(0,0)$ of system $(9)$ is locally asymptotically stable. If
$\beta f'(0)-\alpha_0^2>0$, then as $\alpha$ passes through the
critical value $\alpha_0=af'(0)-1$, there is a Hopf bifurcation at
the equilibrium $(0,0)$.

\subsection{Stability of bifurcating periodic solutions}

In this subsection, the stability of the bifurcating periodic
solutions is studied. The method used is based on the normal form
theory [Hassard et al., 1981]. For notational convenience, let the
above $\alpha=\alpha_0+\gamma$, so $\gamma = 0$ is the Hopf
bifurcation value for system (9).

The eigenvactor $v_1$ associated with $\lambda_1=\mu+i\omega$ is
\begin{equation*}
v_1=\left[\begin{array}{c}1 \\ \mu+i\omega-af'(0)+1 \end{array}
\right]
\end{equation*}
Define
\begin{equation*}
P=(\mbox{Re}\,v_1,\,
-\mbox{Im}\,v_1)=\left[\begin{array}{cc}1&0\\
\mu-af'(0)+1&-\omega
\end{array}\right]
\end{equation*}
and
\begin{equation*}
\left[\begin{array}{c}y_1\\ y_2 \end{array}
\right]=P^{-1}\left[\begin{array}{c}x \\ y\end{array}\right]
\end{equation*}
Then, the system of $y_1,\,y_2$ is obtained from (10) as
\begin{equation*}
\begin{array}{l}
\dot{y}_1=\mu y_1-\omega y_2+\frac{1}{2}af''(0)y_1^2
+\frac{1}{6}af'''(0)y_1^3+\cdots\\
\dot{y}_2=\omega y_1+\mu y_2
+\frac{a(\mu-af'(0)+1)+\beta}{2\omega}f''(0)y_1^2
+\frac{a(\mu-af'(0)+1)+\beta}{6\omega}f'''(0)y_1^3+\cdots
\end{array}
\end{equation*}

Since one only needs to evaluate $\mu_2,\tau_2,\mbox{and}\,
\beta_2$, which are obtained from $C_1(0)$ alone [Hassard et al.,
1981], one may set
\begin{equation*}
\alpha=\alpha_0=af'(0)-1
\end{equation*}
so that, in the above system of $y_1$ and $y_2$,
\begin{equation*}
\mu=0,\mbox{  and }\, \omega=\omega_0=\sqrt{\beta
f'(0)-(af'(0)-1)^2}
\end{equation*}
Thus, the system becomes
\begin{equation*}
\begin{array}{l}
\dot{y}_1=-\omega_0y_2+F_1(y_1,y_2;0)\\
\dot{y}_2=\omega_0y_1+F_2(y_1,y_2;0)
\end{array}
\end{equation*}
where
\begin{equation*}
F_1(y_1,y_2;0)=\frac{1}{2}af''(0)y_1^2+\frac{1}{6}af'''(0)y_1^3+\cdots
\end{equation*}
and
\begin{equation*}
F_2(y_1,y_2;0)=\frac{a(1-af'(0))+\beta}{2\omega_0}f''(0)y_1^2
+\frac{a(1-af'(0))+\beta}{6\omega_0}f'''(0)y_1^3+\cdots
\end{equation*}

Next, the following quantities are evaluated at $\gamma=0\,
(\mbox{i.e.}\,\alpha=\alpha_0)\,\mbox{and}\,(y_1,y_2)=(0,\,0)$.
One has
\begin{equation}
\begin{array}{rl}
g_{11}=&\frac{1}{4}\left[\frac{\partial^2F_1}{\partial y_1^2}
+\frac{\partial^2F_1}{\partial y_2^2}
+i\left[\frac{\partial^2F_2}{\partial y_1^2}
+\frac{\partial^2F_2}{\partial y_2^2}\right]\right] \\
=&\frac{1}{4}f''(0)\left[a+i\frac{a(1-af'(0))+\beta}{\omega_0}\right]\\
g_{02}=& \frac{1}{4}\left[\frac{\partial^2F_1}{\partial y_1^2}-
\frac{\partial^2F_1}{\partial y_2^2}
-2\frac{\partial^2F_2}{\partial y_1 \partial y_2}
+i\left[\frac{\partial^2F_2}{\partial y_1^2}
-\frac{\partial^2F_2}{\partial y_2^2}
+2\frac{\partial^2F_1}{\partial y_1 \partial y_2}\right]\right]\\
=&\frac{1}{4}f''(0)\left[a+i\frac{a(1-af'(0))+\beta}{\omega_0}\right]\\
g_{20}=&\frac{1}{4}\left[\frac{\partial^2F_1}{\partial y_1^2}-
\frac{\partial^2F_1}{\partial y_2^2}
+2\frac{\partial^2F_2}{\partial y_1 \partial y_2}
+i\left[\frac{\partial^2F_2}{\partial y_1^2}
-\frac{\partial^2F_2}{\partial y_2^2}
-2\frac{\partial^2F_1}{\partial y_1 \partial y_2}\right]\right]\\
=&\frac{1}{4}f''(0)\left[a+i\frac{a(1-af'(0))+\beta}{\omega_0}\right]\\
g_{21}=&\frac{1}{8}\left[\frac{\partial^3F_1}{\partial y_1^3}
+\frac{\partial^3F_1}{\partial y_1 \partial y_2^2}
+\frac{\partial^3F_2}{\partial y_1^2 \partial y_2}
+\frac{\partial^3F_2}{\partial y_2^3} +i\left[
\frac{\partial^3F_2}{\partial y_1^3}
+\frac{\partial^3F_2}{\partial y_1 \partial y_2^2}
-\frac{\partial^3F_1}{\partial y_1^2 \partial y_2}
-\frac{\partial^3F_1}{\partial y_2^3}\right]\right]\\
=&\frac{1}{8}f'''(0)\left[a+i\frac{a(1-af'(0))+\beta}{\omega_0}\right]
\end{array}
\end{equation}

Based on the above analysis, one can see that each $g_{ij}$ in
(11) is determined by the parameters in (10). Thus, one can
compute the following quantities:
\begin{equation}
\begin{array}{rl}
C_1(0)=&\frac{i}{2\omega_0}\left[g_{20}g_{11}-2|g_{11}|^2
-\frac{1}{3}|g_{02}|^2\right]+\frac{g_{21}}{2}\\
\mu_2=&-\mbox{Re}\,C_1(0)/\mu'(0)=2\mbox{Re}\,C_1(0)\\
\tau_2=&-(\mbox{Im}\,C_1(0)+\mu_2\omega'(0))/\omega_0 \\
\beta_2=&2\mbox{Re}\,C_1(0)
\end{array}
\end{equation}

Now, the main results of this subsection are summarized as
follows.
\medskip

{\bf THEOREM 2:} In formulas (12), $\mu_2$ determines the
direction of the Hopf bifurcation: if $\mu_2>0\,\,(<0)$, then the
Hopf bifurcation is supercritical (subcritical) and the
bifurcating periodic solutions exist for $\alpha > \alpha_0\,\,
(<\alpha_0)$; $\beta_2$ determines the stability of the
bifurcating periodic solutions: the solutions are orbitally stable
(unstable) if $\beta_2 <0 \,\,(>0)$; and $\tau_2$ determines the
period of the bifurcating periodic solutions: the period increases
(decreases) if $\tau_2 >0\, \,(< 0)$.

\subsection{A numerical example}

In this subsection, an example in the form of system (9) is
discussed, with $a=1.6, f(\cdot)=\mbox{tanh}(\cdot)\,
\mbox{and}\,\beta=0.5$.

By Theorem 1, one can determine that
\begin{equation*}
\alpha_0=0.6, \hspace{1cm} \omega_0=0.3742
\end{equation*}
It follows from the results in Subsection 3.2 that
\begin{equation*}
\mu_2=-0.4, \hspace{1cm} \tau_2=0.9446, \hspace{1cm} \beta_2=-0.4
\end{equation*}
These calculations prove that the equilibrium $(0,0)$ is stable
when $\alpha>\alpha_0$, as shown by Fig. 1, where $\alpha=0.7$.
When $\alpha$ passes through the critical value $\alpha_0=0.6$,
the equilibrium losses its stability and a Hopf bifurcation
occurs, i.e., a family of periodic solutions bifurcate out of the
equilibrium. Each individual periodic orbit is stable since
$\beta_2<0$. Since $\mu_2<0$, the bifurcating periodic solutions
exist at least for values of $\alpha$ slightly less than the
critical value. Choosing $\alpha=0.02$, as predicted by the
theory, Fig. 2 shows that there indeed is a stable limit cycle.
Since $\tau_2>0$, the period of the periodic solutions increases
as $\alpha$ increases. For $\alpha=0.5$, the phase plot and the
waveform plot are shown in Fig. 3. Comparing Fig. 2 with Fig. 3,
one can conclude that the period of $\alpha=0.5$ is longer than
that of $\alpha=0.02$.

\begin{figure}[htb]
\centering
\includegraphics[width=13cm]{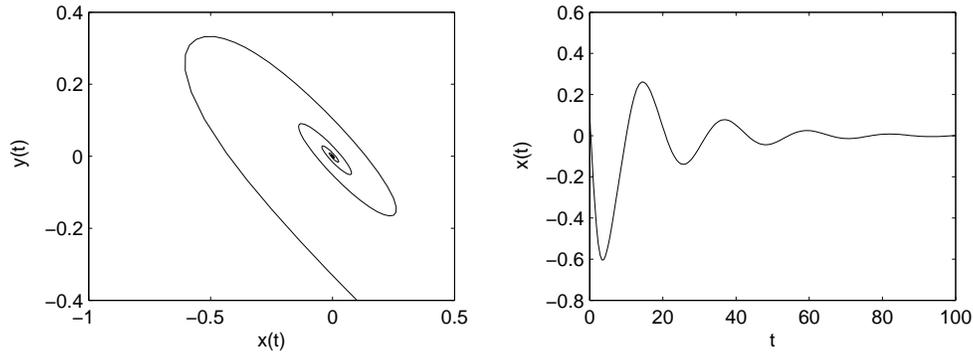}
\caption{Phase plot and waveform plot for system (9) with
$\alpha=0.7$}
\end{figure}
\begin{figure}[htb]
\centering
\includegraphics[width=13cm]{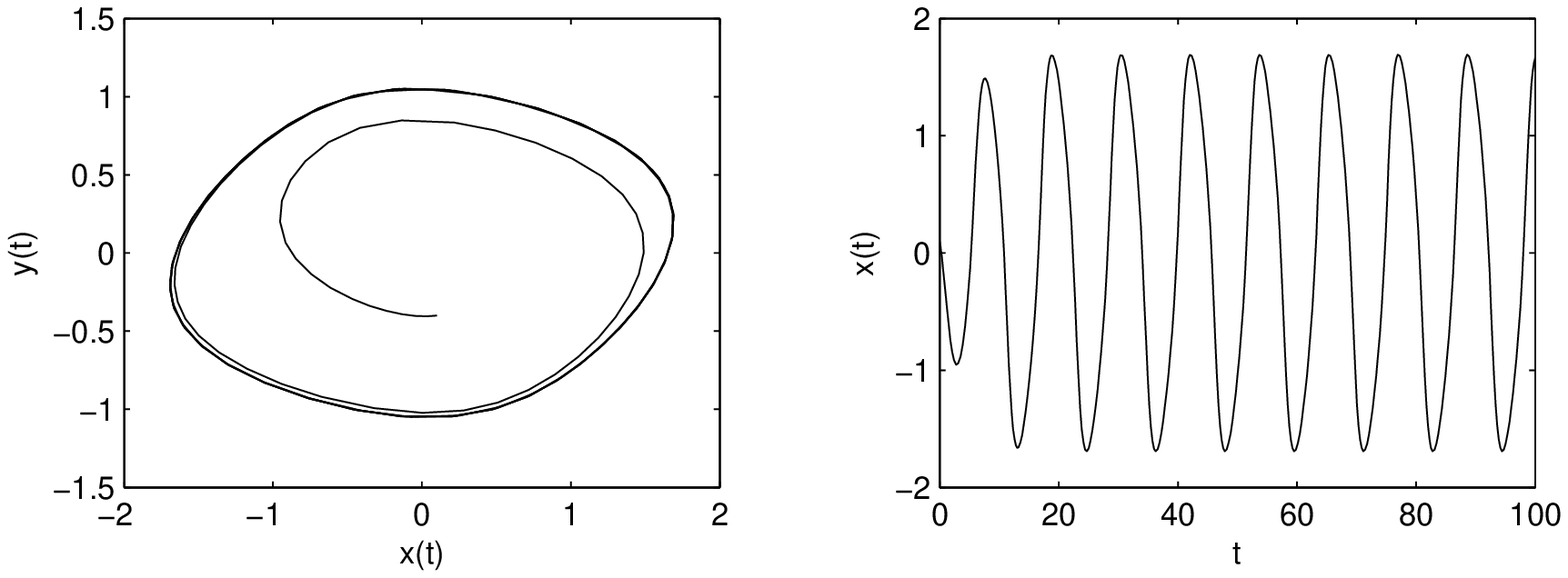}
\caption{Phase plot and waveform plot for system (9) with
$\alpha=0.02$}
\end{figure}
\begin{figure}[htb]
\centering
\includegraphics[width=13cm]{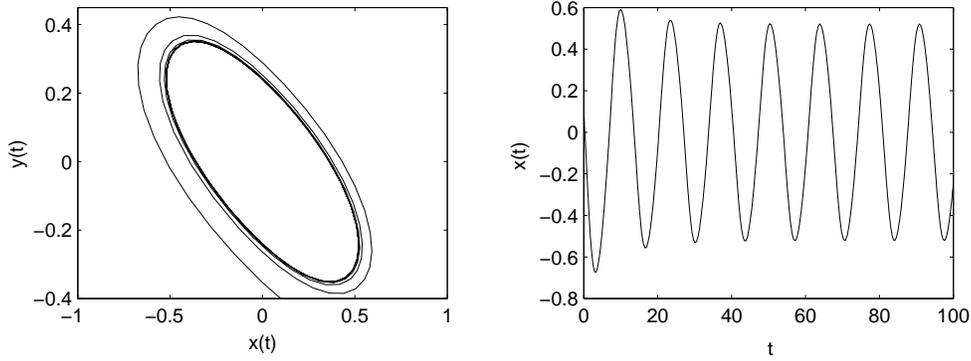}
\caption{Phase plot and waveform plot for system (9) with
$\alpha=0.5$}
\end{figure}

\section{Chaos in an excited tabu learning single neuron model}

In this section, the dynamical behaviors of a tabu learning single
neuron model, with external sinusoidal input, are studied.

In system (8), let $I$ be
\begin{equation*}
I(t)=\epsilon \sin(\omega t)
\end{equation*}
and let the nonlinear activation function be
\begin{equation}
f(x)=pxe^{-(px)^2/\sigma^2}
\end{equation}
where $\epsilon,\,\omega,\,p$, and $\sigma$ are positive
constants. Also, in system (8), set the parameters be $C=1,\,
R=7,\,a=0.3,\,\epsilon=0.2,\,\omega=2\pi,\,\alpha=0.6,\,
\beta=0.9$, and set $p=8,\,\sigma^2=0.2$ in (13). It should be
noted that several other parameters have also been examined,
showing similar dynamical phenomena. Due to the space limitation,
those results are not presented here.

The phase plot of $x$ and $y$ and the waveform plot of $x$ are
shown in Fig. 4 and Fig. 5, respectively. And in Fig. 6, the power
spectrum of the neuron state $x$, calculated using the general
FFT, is plotted.

\begin{figure}
\begin{minipage}[htb]{0.55\linewidth}
\centering
\includegraphics[width=7cm]{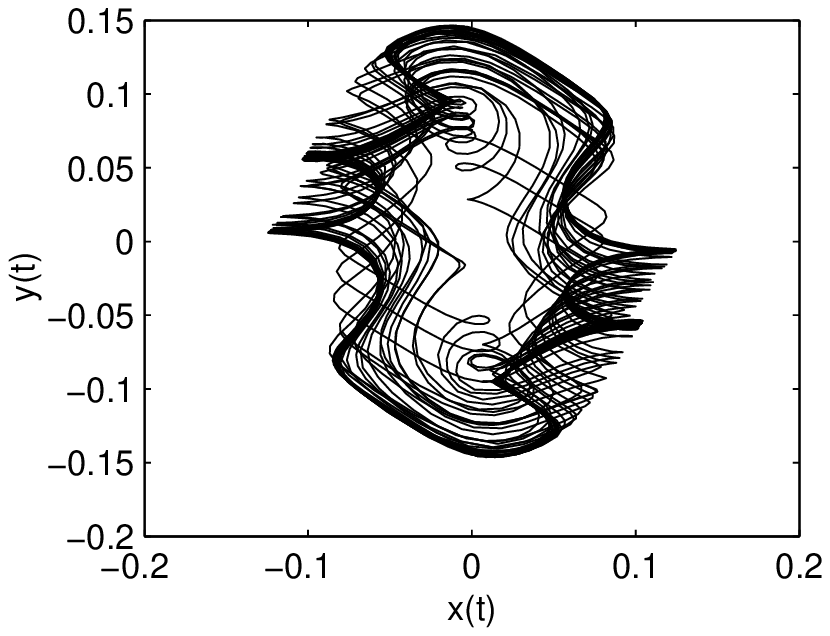}
\caption{Phase plot}
\end{minipage}%
\hspace{-1.2cm}%
\begin{minipage}[htb]{0.5\linewidth}
\centering
\includegraphics[width=7cm]{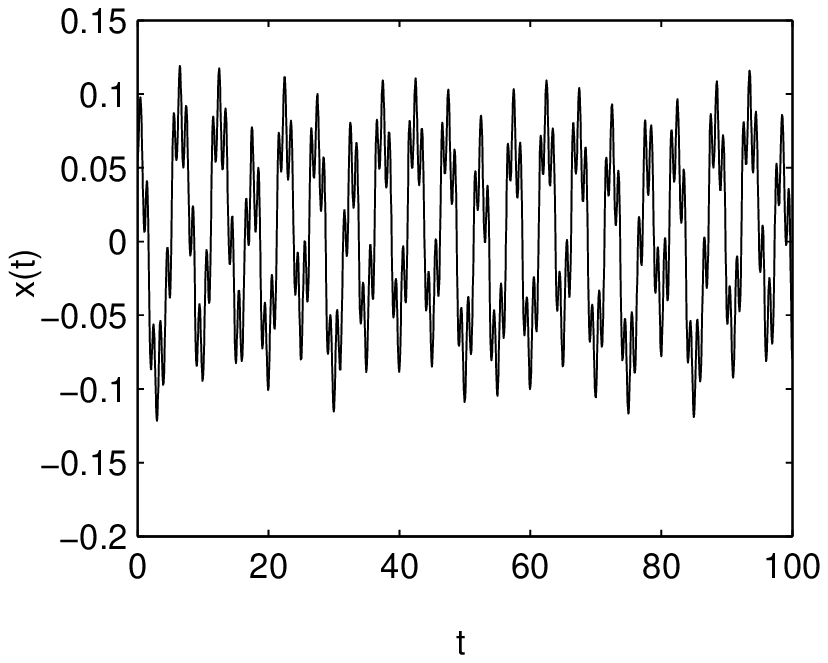}
\caption{Waveform plot of $x$}
\end{minipage}
\end{figure}
\begin{figure}[htb]
\centering
\includegraphics[width=7cm]{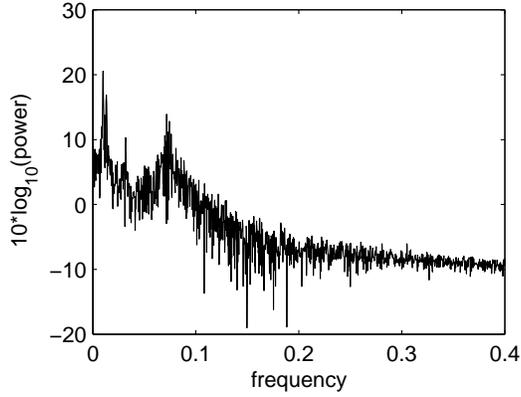}
\caption{Power spectrum of the time series of $x$}
\end{figure}

In order to calculate the largest Lyapunov exponent, the method
introduced in [Wolf et al., 1985] is used, in which one monitors
the long-term evolution of a single pair of nearby orbits. The
largest Lyapunov exponent is defined as
\begin{equation*}
\lambda=\frac{1}{t_M-t_0}\sum_{k=1}^M \mbox{log}_2
\frac{L'(t_k)}{L(t_{k-1})}
\end{equation*}
where $M$ is the total number of replacement steps, $L(t_{k-1})$
is the distance between the two initial points at the time instant
$t_{k-1}$. After a time step $\delta=t_k-t_{k-1}$, the initial
length will have evolved to length $L'(t_k)$, as detailed in [Wolf
et al., 1985]. In the simulations, the largest Lyapunov exponent
$\lambda$ is calculated from a time series of $N=100,000$ points.
The largest Lyapunov exponent is obtained as $\lambda=0.7030$.

From Figs. 4-6 and the largest Lyapunov exponent, one can see that
there exists chaotic behavior in this nonautonomous tabu learning
single neuron model.

\section{Chaos in a two-neuron model with a linear proximity function}

Now, consider a two-neuron model with a linear proximity function.
From (6) and (7), one knows that there are totally four
differential equations in this system:
\begin{equation}
\begin{array}{rl}
C_1\dot{u}_1=&-\frac{1}{R_1}u_1+T_{11}V_1+T_{12}V_2+I_1+J_1\\
C_2\dot{u}_2=&-\frac{1}{R_2}u_2+T_{21}V_1+T_{22}V_2+I_2+J_2\\
\dot{J}_1=&-\alpha J_1-\beta V_1 \\
\dot{J}_2=&-\alpha J_2-\beta V_2
\end{array}
\end{equation}

In the following simulations, without loss of generality, let
$C_1=C_2=1$, $R_1=R_2=10$, $I_1=I_2=0$, $\alpha=0.1$, $\beta$ be a
variable parameter, and let the weight matrix be
\begin{equation*}
T=\left[\begin{array}{cc}0.1&0.5\\-1&2\end{array}\right]
\end{equation*}
Moreover, let the activation function be $V_i=f(u_i)
=\mbox{tanh}(5u_i)$. It should be noted that several other
parameters have also been examined, showing similar dynamical
behaviors. Due to space limitation, those results are not
presented here.

\begin{figure}[htb]
\centering
\includegraphics[width=15cm]{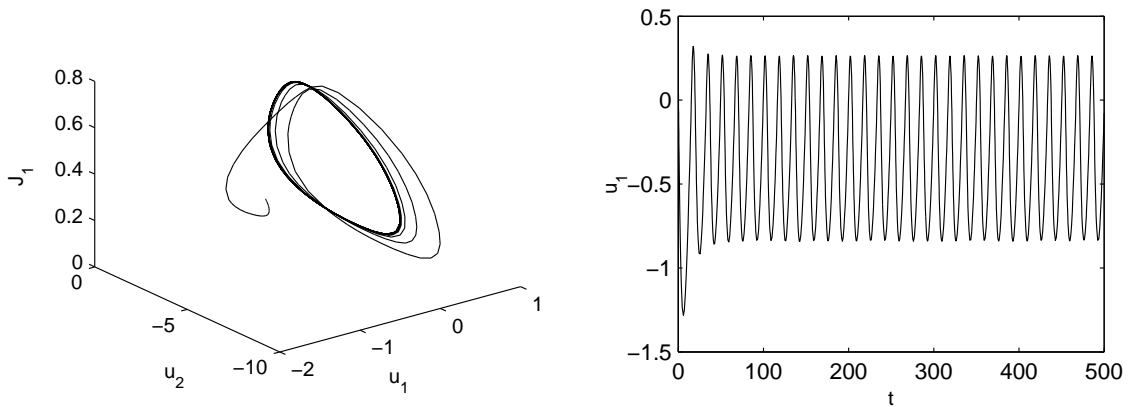}
\caption{Phase plot and waveform plot of the two-neuron model (14)
($\beta=0.1$)}
\end{figure}
\begin{figure}[htb]
\centering
\includegraphics[width=15cm]{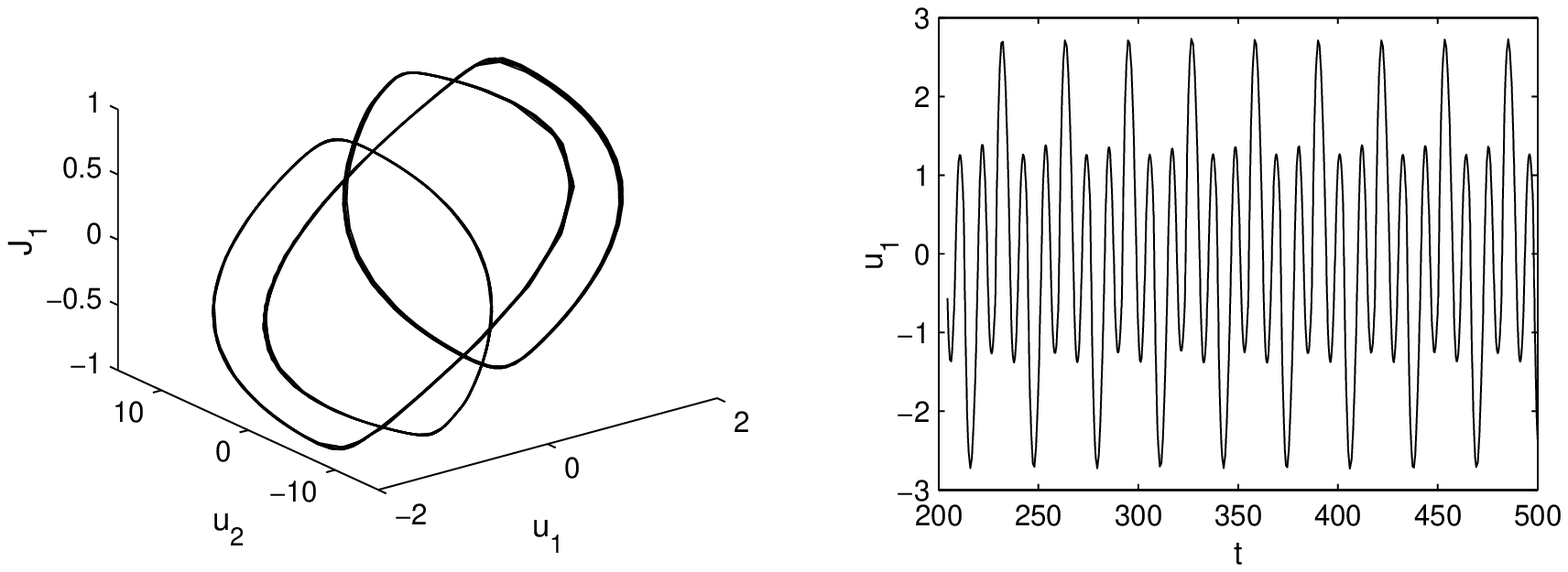}
\caption{Phase plot and waveform plot of the two-neuron model (14)
($\beta=0.5$)}
\end{figure}
\begin{figure}[htb]
\centering
\includegraphics[width=15cm]{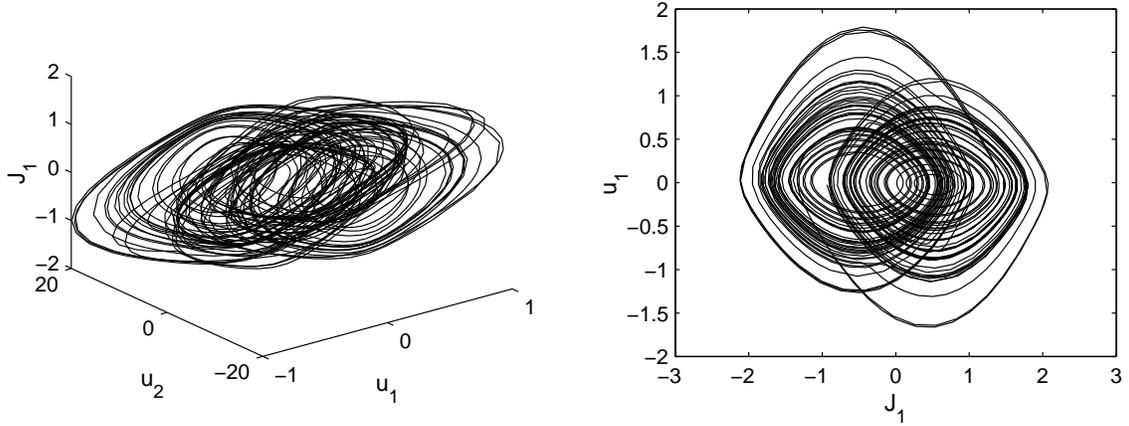}
\caption{3-D and 2-D phase plots of the two-neuron model (14)
($\beta=1$)}
\end{figure}
\begin{figure}[htb]
\centering
\includegraphics[width=15cm]{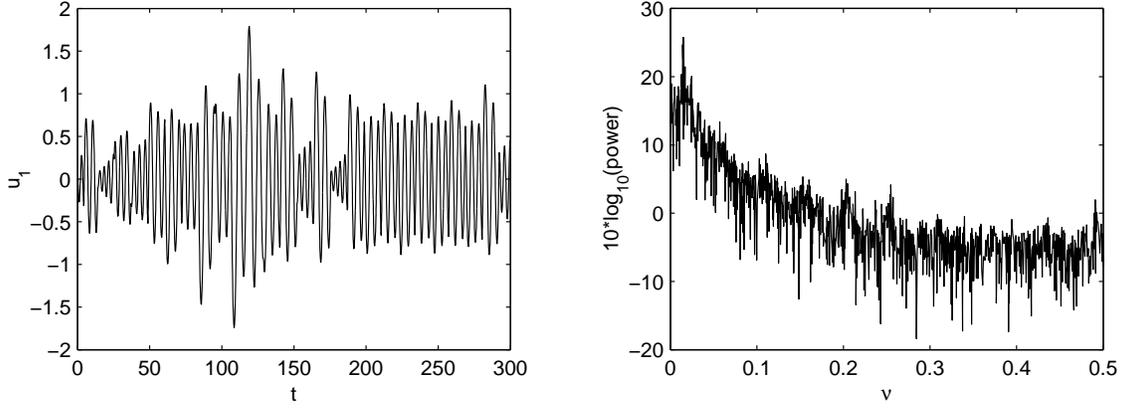}
\caption{Waveform plot and power spectrum of the two-neuron model
(14) ($\beta=1$)}
\end{figure}

Start from the case of $\beta=0.1$, with which the neural network
produces periodic solutions. Fig. 7 shows the 3-D phase plot and
the waveform plot of $u_1$. Fig. 8 shows the 3-D phase plot and
the waveform plot of the network orbit when $\beta=0.5$. In this
case, it is also a periodic solution, but it is different from
that of the case $\beta=0.1$. When $\beta=1$, the orbit of the
neural network is chaotic, and the largest Lyapunov exponent is
0.2419. The 3-D and 2-D plots of the chaotic attractor are shown
in Fig. 9. The corresponding waveform and power spectrum are shown
in Fig. 10, where the spectrum has been truncated at $\nu$=0.5 HZ.

\section{Chaos in a two-neuron model with a quadratic proximity function}

If the quadratic proximity function (5) is selected, then the tabu
learning neural network that performs gradient descent
minimization on $E_t$ has the following state equations:
\begin{equation}
\begin{array}{lcl}
C_i\dot{u}_i&=&-\frac{1}{R_i}u_i+\sum_jT_{ij}V_j+I_i
-\frac{\partial{F_t}(V)}{\partial{V_i}}\\
&=&-\frac{1}{R_i}u_i+\sum_j(T_{ij}+S_{ij}(t))V_j+I_i+J_i(t)
\end{array}
\end{equation}
where
\begin{equation*}
\begin{array}{rl}
S_{ij}(t):=&-\beta\int_0^te^{\alpha(s-t)}V_i(s)V_j(s)\,ds
\hspace{0.5cm} (i\neq j)\\
S_{ii}(t):=&0\\
J_i(t):=& -\beta(n-1)\int_0^te^{\alpha(s-t)}V_i(s)\,ds
\end{array}
\end{equation*}
in which $n$ is the number of neurons. Therefore, $S_{ij}(t),\,
i\neq j$, and $J_i(t)$ satisfy the following learning equations:
\begin{equation}
\dot{S}_{ij}=-\alpha S_{ij}-\beta V_iV_j
\end{equation}
\begin{equation}
\dot{J_i}=-\alpha J_i-\beta (n-1)V_i
\end{equation}

There are totally six differential equations in this model:
\begin{equation}
\begin{array}{rl}
C_1\dot{u}_1=&-\frac{1}{R_1}u_1+(T_{12}+S_{12}(t))V_2+I_1+J_1(t)\\
C_2\dot{u}_2=&-\frac{1}{R_2}u_2+(T_{21}+S_{21}(t))V_1+I_2+J_2(t)\\
\dot{S}_{12}=&-\alpha S_{12}-\beta V_1V_2 \\
\dot{S}_{21}=&-\alpha S_{21}-\beta V_2V_1 \\
\dot{J}_1=&-\alpha J_1-\beta V_1\\
\dot{J}_2=&-\alpha J_2-\beta V_2
\end{array}
\end{equation}
For this model, only its chaotic behavior is discussed. Let
$C_1=C_2=1$, $R_1=R_2=10$, $I_1=I_2=0$, $\alpha=0.1$, $\beta=100$,
$V_i=\mbox{tanh}(10u_i)$, and the weight matrix be
\begin{equation*}
T=\left[\begin{array}{cc}0&30\\50&0\end{array}\right]
\end{equation*}
Then, the neural network is chaotic. The 3-D and 2-D phase plots
of the chaotic attractor are shown in Fig. 11 and Fig. 12,
respectively. The waveform plot and power spectrum of $u_1$ are
shown in Fig. 13, in which the largest Lyapunov exponent is
0.3160.

\begin{figure}[htb]
\centering
\includegraphics[width=12cm]{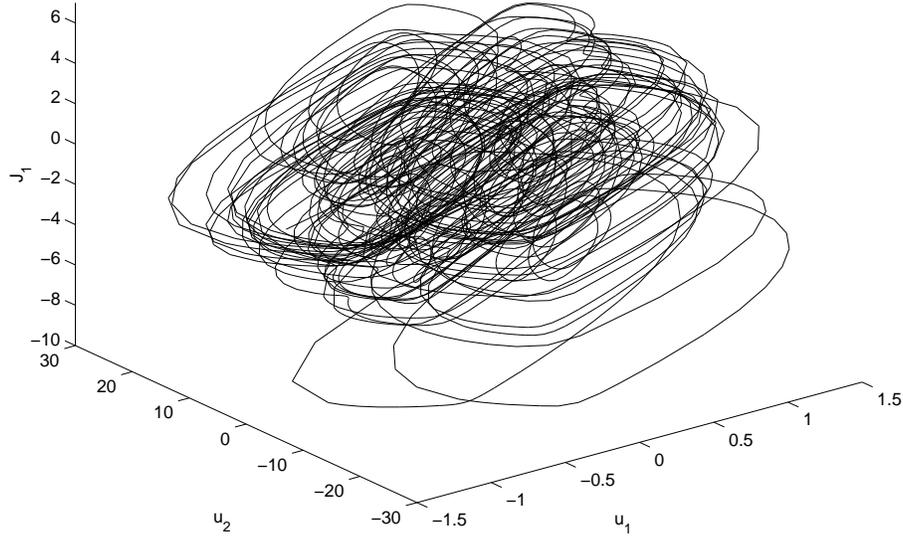}
\caption{3-D phase plot of the chaotic attractor of the two-neuron
model (18)}
\end{figure}
\begin{figure}[htb]
\centering
\includegraphics[width=12cm]{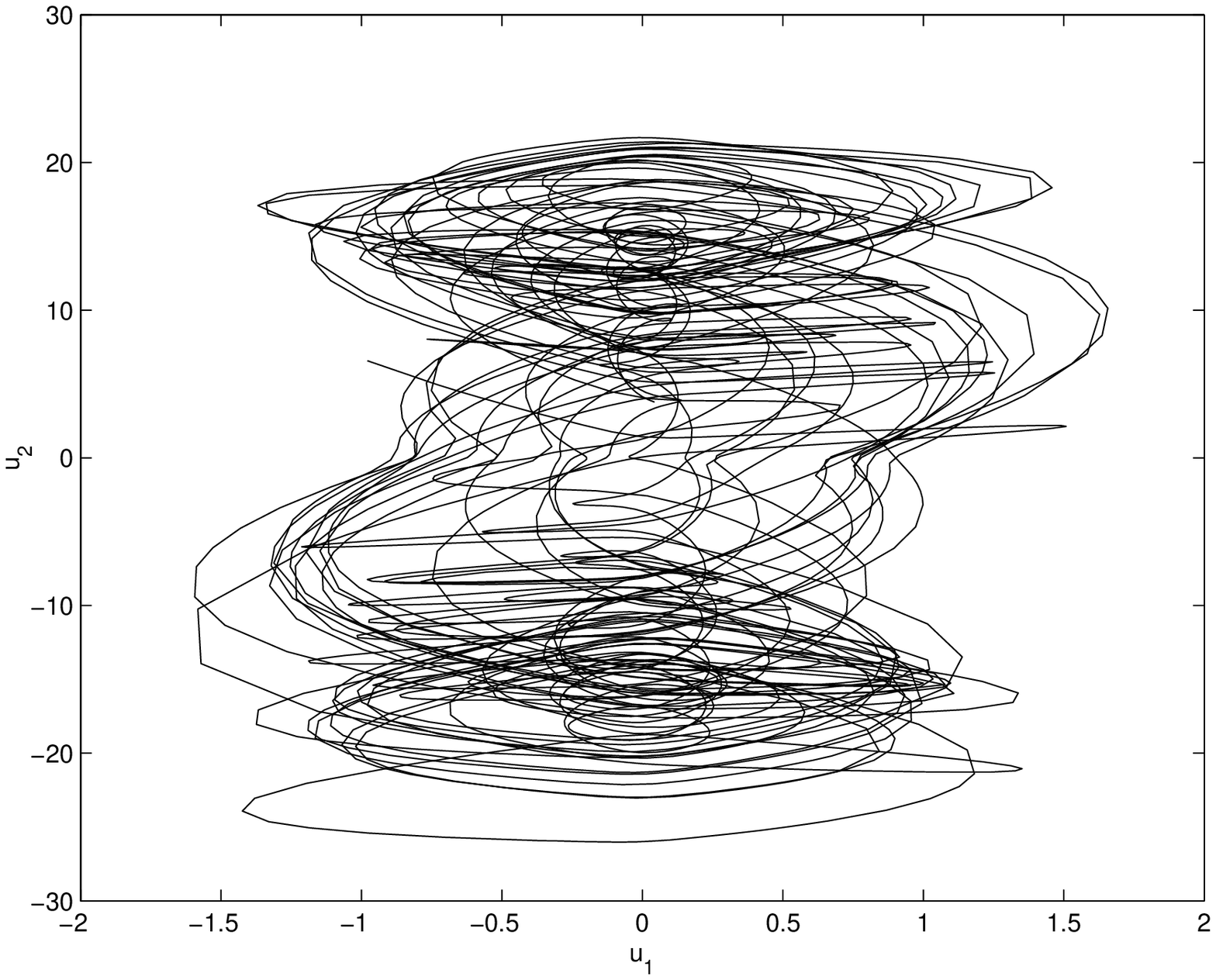}
\caption{2-D phase plot of the chaotic attractor of the two-neuron
model (18)}
\end{figure}
\begin{figure}[htb]
\centering
\includegraphics[width=15cm]{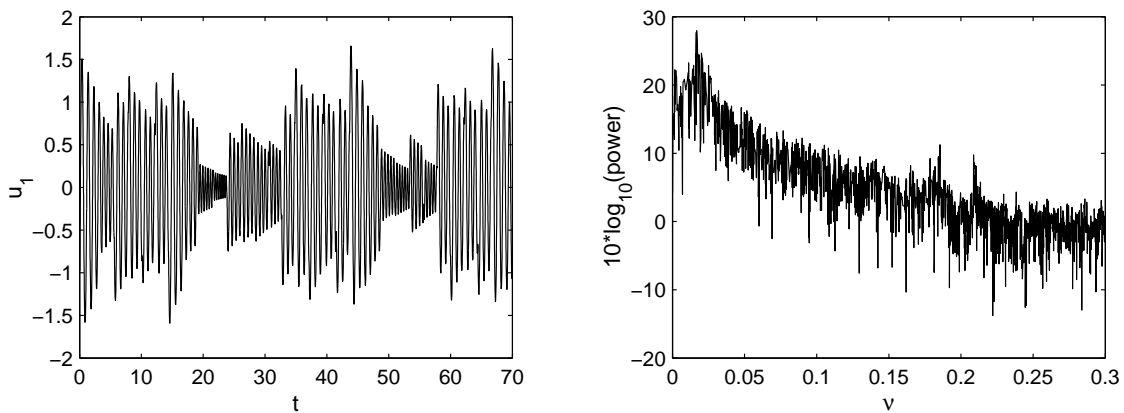}
\caption{Waveform plot and power spectrum plot of the two-neuron
model (18)}
\end{figure}

\section{Conclusions}

The nonlinear dynamical behaviors of some tabu learning neuron
models have been studied in this paper. By choosing the memory
decay rate as a bifurcation parameter, it has been proved that
Hopf bifurcation occurs in the single neuron model. The stability
of bifurcating periodic solutions and the direction of the Hopf
bifurcation are determined based on the normal form theory. From
the waveform diagrams, the phase plots, the power spectra, and the
largest Lyapunov exponents, one can find chaotic phenomena in the
single neuron model with small external sinusoidal input and in
the two-neuron models. These tabu learning models, although
simple, have rich complex dynamics, and deserve further
investigations.

\subsection*{Acknowledgements}

This work was supported by the National Natural Science Foundation
of China under Grant 60271019, the Youth Science and Technology
Foundation of UESTC under Grant YF020207, and the Hong Kong
Research Grants Council under the CERG Grant CityU 1115/03E.

\section*{References}
\begin{description}
\item Babcock, K.L. \& Westervelt, R.M. [1987] ``Dynamics of
simple electronic neural networks'', {\it Physica} {\bf D 28},
305-316.

\item Beyer, D.A. \& Ogier, R.G. [1991] ``Tabu learning: A neural
network search method for solving nonconvex optimization
problems", {\it Proc. of the IJCNN} (Singapore), 953-961.

\item Bondarenko, V.E. [1997] ``High-dimensional chaotic neural
network under external sinusoidal force", {\it Phys. Lett.} {\bf A
236}, 513-519.

\item Das, A., Das, P. \& Roy, A.B. [2002] ``Chaos in a
three-dimensional general model of neural network", {\it Int. J.
Bifurcation and Chaos} {\bf 12}, 2271-2281.

\item Das II, P.K., Schieve, W.C., Zeng, Z.J. [1991] ``Chaos in an
effective four-neuron neural network", {\it Phys. Lett.} {\bf A
161}, 60-66.

\item Gilli, M. [1993] ``Strange attractors in delayed cellular
neural networks", {\it IEEE Trans. Circ. Sys. -- I}, {\bf 40},
849-853.

\item Glover, F. [1989] ``Tabu search, part I", {\it ORSA J.
Comput.} {\bf 1}, 190-206.

\item Glover, F. [1990] ``Tabu search, part II", {\it ORSA J.
Comput.} {\bf 2}, 4-32.

\item Gopalsamy, K., Leung, I. \& Liu, P. [1998] ``Global
Hopf-bifurcation in a neural netlet", {\it Appl. Math. Comput.}
{\bf 94}, 171-192.

\item Hassard, B. D., Kazarinoff, N.D., \& Wan, Y.H. [1981], {\it
Theory and Applications of Hopf Bifurcation} (Cambridge University
Press, Cambridge).

\item Kepler, T.B., Datt, S., Meyer, R.B., Abbott, L.F. [1990]
``Chaos in a neural network circuit", {\it Physica} {\bf D 46},
449-457.

\item Kurten, K.E. \& Clark, J.W. [1986] ``Chaos in neural
systems'', {\it Phys. Lett.} {\bf A 144}, 413-418.

\item Li, C., Liao, X. \& Yu, J. [2003] ``Generating chaos by
Oja's rule", {\it Neurocomputing} {\bf 55}, 731-738.

\item Li, C., Yu, J., \& Liao, X. [2001] ``Chaos in a three-neuron
hysteresis Hopfield-type neural network", {\it Phys. Lett.} {\bf A
285}, 368-372.

\item Liao, X., Wong, K.-W., Leung, C.-S. \& Wu, Z. [2001] ``Hopf
bifurcation and chaos in a single delayed neuron equation with
non-monotonic activation function", {\it Chaos, Solitons and
Fractals} {\bf 12}, 1535-1547.

\item Liao, X., Wong, K.-W. \& Wu, Z. [2001] ``Bifurcation
analysis on a two-neuron system with distributed delays", {\it
Physica} {\bf D 149}, 123-141.

\item Liao, X. Wu, Z. \& Yu, J. [1999] ``Stability switches and
bifurcation analysis of a neural network with continuous delay",
{\it IEEE Trans. Sys. Man Cybern. -- A} {\bf 29}, 692-696.

\item Lu, H. [2002] ``Chaotic attractors in delayed neural
networks", {\it Phys. Lett.} {\bf A 298}, 109-116.

\item Olien, L. \& Belair, J. [1997] ``Bifurcations, stability and
monotonicity properties of a delayed neural network model", {\it
Physica} {\bf D 102}, 349-363.

\item Ueta, T. \& Chen, G. [2001] ``Chaos and bifurcation in
coupled networks and their control", in {\it Controlling Chaos and
Bifurcations in Engineering Systems}, G. Chen (ed.), (CRC Press,
Boca Raton), 581-601.

\item Wei, J. \& Ruan, S. [1999] ``Stability and bifurcation in a
neural network model with two delays", {\it Physica} {\bf D 130},
255-272.

\item Wolf, A., Swift, J.B., Swinney, H.L., Vastano, J.A. [1985]
``Determining Lyapunov exponents from a time series'', {\it
Physica} {\bf D 16}, 285-317.

\item Zou, F. \& Nossek, J.A. [1993] ``Bifurcation and chaos in
cellular neural network", {\it IEEE Trans. Circ. Sys. -- I} {\bf
40}, 166-172.
\end{description}

\end{document}